\documentclass{an}
\usepackage{graphicx}
\usepackage{times}
\usepackage[comma,authoryear]{natbib}
\usepackage[latin1]{inputenc}

\bibpunct{(}{)}{;}{a}{}{,}

\begin{document}

\Pagespan{789}{}
\Yearpublication{2007}%
\Yearsubmission{2007}%
\Month{11}%
\Volume{999}%
\Issue{88}%

\title{Bayesian approach for g-mode detection, or how to restrict our imagination}

\author{T.~Appourchaux\inst{1}\fnmsep\thanks{Corresponding author:
  \email{Thierry.Appourchaux@ias.u-psud.fr}\newline}
}
\titlerunning{Bayesian approach for helioseismology}
\authorrunning{T.Appourchaux}
\institute{
Institut d'Astrophysique Spatiale, UMR 8617, Universit\'e Paris-Sud, B{\^a}timent 121, 91045 Orsay Cedex, France}

\received{9 October 2007}
\accepted{}
\publonline{later}


\abstract{%
Nowadays, g-mode detection is based upon {\it a priori} theoretical knowledge.  By doing so, detection becomes more restricted to what we can imagine.  {\it De facto}, the universe of possibilities is made narrower.  Such an approach is pertinent for Bayesian statisticians.  Examples of how Bayesian inferences can be applied to spectral analysis and helioseismic power spectra are given.  Our intention is not to give the full statistical framework (much too ambitious) but to provide an appetizer for going further in the direction of a proper Bayesian inference, especially for detecting gravity modes.}

\maketitle

\section{Introduction}
Since the beginning of helioseismology, the detection of g modes has 
been the most challenging quest in our field.  There 
were claims of g-mode detection 
\citep[][]{PDPS83, DT95},  
none of which were confirmed.  Since the 
conception of the SOHO 
mission, one of the goals of this mission was to detect g modes.  
In 1997, following the lack of g-mode detection by SOHO experimenters, the 
Phoebus group was formed, with the aim of detecting g modes.  The 
group set an upper limit to the g-mode amplitude of 10 mm/s at 200 
$\mu$Hz \citep{TA2000}.  Since then, this lower limit has been even decreased
down to about 4.5 mm.s$^{-1}$ for a 10-year observation with a singlet, and down to 1.5 mm.s$^{-1}$ for a multiplet \citep{YE2006}.

Over the years, the Phoebus group has developed several techniques for g-mode detection.  Some of which are based upon having no knowledge of the structure and dynamics of the Sun.  There are also other techniques using {\it a priori} knowledge of the Sun such as using rotational splitting patterns \citep[][and references therein]{YE2006}.  Recent g-mode detection claims rely upon a new detection technique derived from the asymptotic properties of g-mode periods derived from theoretical models \citep{Rafa2007}.  Although these approaches look promising, they are all ba\-sed on what we believe we know about the Sun.  From that point of view, it is time that we turn to an approach that has been talked about a lot: a Bayesian approach to statistical inference.

First, I will present a short introduction to what I believe I understand about a Bayesian approach versus a frequentist approach.  In the second section, I will show simple examples of this can be applied to Fourier analysis.  In the third section, the Bayesian approach is applied to well known example in helioseismology.  I then conclude and give a tentative roadmap for the future.

\section{Bayesians versus Frequentists}
The controversy between Bayesians and Frequentists is related to {\it subjective} versus {\it objective} probabilities.  A frequentist thinks that the laws of physics are {\it deterministic}, while a Bayesian ascribes a belief that the laws of physics are true or {\it operational}.  The {\it subjective} approach to probability was first coined by \citet{Finetti}, the reading of which is extremely enlightening.
For the rest of us, the difference in views can be summarized by this quote from the Wikipedia encyclopedia: {\it Whereas a frequentist and a Bayesian might both assign probability $\frac{1}{2}$ to the event of getting a head when a coin is tossed, only a Bayesian might assign probability $\frac{1}{1000}$ to personal belief in the proposition that there was life on Mars a billion years ago, without intending to assert anything about any relative frequency.}  In short frequentists assign probability to measurable events that can be infinitely measured, while Bayesians assign probability to events that cannot be measured, like the outcome of sport-related bets for instance.  The Bayesian approach is then related to what Reverend Bayes would have understood as {\it the degree of belief}.  The application of Bayes' theorem is then referred to as {\it Bayesian inference}.

\subsection{Bayes' theorem}
The theorem of \citet{Bayes} is known to any kid going to high school.  It relates the probability of an event A given the occurrence of an event B to the probability of the event B given the occurence of the event A, and the probability of occurrence of the event A and B.
\begin{equation}
P({\rm A} | {\rm B}) = \frac{P({\rm B} | {\rm A}) P({\rm A})}{P({\rm B})}
\end{equation}
For example, the probability of having rain given the presence of clouds is related to the probability of having clouds given the presence of rain by Eq. (1).  The term {\it prior probability} is given to $P({\rm A})$ (probability of having rain in general).  The term {\it likelihood} is given to $P({\rm B} | {\rm A})$ (probability of having clouds given the presence of rain) .  The term {\it posterior probability} is given to $P({\rm A} | {\rm B})$ (probability of having rain given the presence of clouds).  The term {\it normalization constant} is given to $P({\rm B})$ (probability of having clouds in general).  

This theorem can then be transposed to anything that we know, or any information known {\it a priori}:
\begin{equation}
P(\Omega | {\rm D,I}) = \frac{P(\Omega | {\rm I})P({\rm D} | \Omega, {\rm I})}{P({\rm D} | {\rm I})}
\end{equation}
where $\Omega$ are the observables for which we seek the {\it posterior probability}, {\rm D} is the observed data set, and {\rm I} is the information.  The {\it prior probability} of the observables is given by $P(\Omega | {\rm I})$: this is the way to quantify our {\it belief} about what we seek.  

\subsection{Nuisance parameters}
In theory, it looks quite simple but in practice the derivation of Eq (2) could be somewhat complicated.  This is especially difficult when there are a subset of observables ($\Omega_{b}$) that are not known and need to be eliminated.  In that case, it is required to integrate (or marginalize) over the {\it unwanted } parameters as follows:
\begin{equation}
P(\Omega_{a} | {\rm D,I}) = \int_{\Omega_{b}} P(\Omega_{a} , \Omega_{b} | {\rm D,I}) {\rm d} \Omega_{b}
\end{equation}
Replacing Eq. (2) in Eq. (3), we get:
\begin{equation}
P(\Omega_{a} | {\rm D,I}) = \int_{\Omega_{b}} \frac{P(\Omega_a, \Omega_b | {\rm I}) P({\rm D} | \Omega_a, \Omega_b, {\rm I}) }{P({\rm D} | {\rm I})}
 {\rm d} \Omega_{b}
 \end{equation}
with the {\it prior} probability expressed using the product rule as:
\begin{equation}
 P(\Omega_a, \Omega_b | {\rm I})=P(\Omega_b | \Omega_a, {\rm I})P(\Omega_a | {\rm I})
\end{equation}
Equation (4) can then be integrated provided that the {\it prior probability} is assigned. 

\subsection{Role of the prior}
If the parameters $\Omega_a$ and $\Omega_b$ are supposed not to be correlated, the {\it prior} probability $P(\Omega_a, \Omega_b | {\rm I})$ (Eq. 5) can then simply be expressed as:
\begin{equation}
P(\Omega_a, \Omega_b | {\rm I})=P(\Omega_a | {\rm I})P(\Omega_b | {\rm I})
\end{equation}
The {\it prior probability} will then express what we believe we know (or not) about the parameters.  The most obvious {\it prior probability} is the one that is uniformly distributed over some range of the parameters $\Omega_b$ of interest.  The choice of the prior is related to the amount of information at our disposal.  The role of the prior and its impact on the {\it posterior probability} should ideally be as small as possible.  \citet{Sivia} show the impact of various priors on the outcome of a Bayesian analysis.  The ideas developed in this article are also used by Thierry Toutain (private communication) for inferring high frequency p-mode splitting hampered by mode blending \citep{TA2000a}.  

This discussion of the role of the prior is beyond the scope of this article.  Here I am just touching the tip of the iceberg.  I encourage the reader to seek other sources of information \citep{Jaynes}.

\subsection{Parameters estimation and error bars}
As soon as the {\it posterior probability} is known, we can derive an estimate of the parameter using:
\begin{equation}
<\Omega_a> = \int \Omega_a P(\Omega_{a} | {\rm D,I}) {\rm d}\Omega_a
\end{equation}
with the following rms error:
\begin{equation}
\sigma_{\Omega_a} = \int (\Omega_a-<\Omega_a>)^ 2 P(\Omega_{a} | {\rm D,I}) {\rm d}\Omega_a
\label{rms}
\end{equation}
This latter expression is used in the next section for the case of spectral analysis.

\section{Spectral analysis revisited}
\citet{Bretthorst}, using a Bayesian approach to the analysis of the time series of a pure sine wave embedded in noise having a gaussian distribution, demonstrated that the {\it posterior probability} for the angular frequency $\omega$ of the sine wave can be written as:
\begin{equation}
P(\omega | {\rm D}, \sigma, {\rm I}) \propto e^{\frac{C(\omega)}{\sigma^2}}  
\end{equation}
where {\rm D} are the data ($d_i$ taken at time $t_i$), $\sigma$ is the rms value of the noise assumed to be known, and $C(\omega)$ is the so-called Schuster periodogram given by:
\begin{equation}
C(\omega)=\frac{1}{N} |\sum_{i=1}^{N} d_i e^{j\omega t_i} |^2
\end{equation}
This periodogram that is today called the Dis\-crete Fou\-rier Transform was first derived by \cite{Schuster}.  Following this approach, \citet{Jaynes} demonstrated using Eq.~(\ref{rms}) that for a pure sine wave of amplitude $A$,  the rms error on the frequency $\nu=\omega/2\pi$ is given by:
\begin{equation}
\delta \nu = \frac{\sqrt{6}}{\pi} \frac{\sigma}{A\sqrt{N}}\frac{1}{T}
\end{equation}
where $T$ is the observing time and $N$ is the number of samples taken.
This formula is the same as given by \citet{Cuypers1987} and derived by \citet{Koen1999}.  This equation shows that the Rayleigh criterion ($1/T$) for frequency resolution is very pessimistic compared to the precision with which frequencies can be measured.  This feature provided by pure sine wave will be fully used by the CoRoT Data Analysis Team for classical stellar pulsators as outlined by \citet{Appourchaux2007}.

This revisitation of spectral analysis by Bretthorst is extremely useful when one wants to understand how to apply a Bayesian analysis to helioseismology.  

\section{Bayesian inference for astero- and helioseismology}
Bayesian inference has recently been used in asteroseismology by \citet{Brewer} and applied to several stars \citep{Carrier2007, Bedding2007}.  Unfortunately, the assumptions about the stochastic nature of the mode excitation are completely ignored in the formulation of the Bayesian inference, even though simulated spectra do integrate the randomness of the excitation.  

The stochastic nature of the mode excitation has been known since \citep{MW84}.  It leads to the formulation by \citet{TDJH86} of the fitting of the p-mode spectrum by Maximum Likelihood Estimation.  In our case, I use this formulation in a very similar manner but instead of expressing the {\it posterior probability} of, say, the frequency of a mode, as a function of the samples taken in time, I will use the samples taken in frequency.  This is because the stochastic nature of the excitation of the harmonic oscillator, representing an eigenmode, prevents us from separating the instrumental noise and the solar noise.  This separation, as we will see later on, is only possible in the Fourier spectrum.  

We apply Eq. (2) to the simple example of a single mode stochastically excited.  In our case, I can then write:
\begin{equation}
P(\Omega_s | {\rm D,I}) = \frac{P(\Omega_s | {\rm I})P({\rm D} | \Omega_s, {\rm I})}{P({\rm D} | {\rm I})}
\end{equation}
with $\Omega_s=(\nu_0, \Gamma, A, B)$, where $\nu_0$ is the mode frequency, $\Gamma$ is the mode linewidth, $A$ is the mode amplitude  and $B$ is the noise.  The {\it prior probability} is given by:
\begin{equation}
P(\Omega_s | {\rm I})=P(\nu_0 | {\rm I})P(\Gamma | {\rm I})P(A | {\rm I})P(B | {\rm I})
\end{equation}
where here I assume that the information about $\nu_0$ has no relation to the information we have about $\Gamma$, $A$ and $B$, and also for all other pairs of parameters.  This is not quite correct for $A$ and $\Gamma$ but this is the choice of the {\it prior} that I made here.  With the assumption of stochastic excitation of the modes, it is known that the power spectrum of the eigenmode is a $\chi^2$ with 2 d.o.f with a mean given essentially by the mode profile plus noise \citet{TDJH86}.  This assumption is sufficient for deriving the {\it likelihood} as:
\begin{equation}
P({\rm D} | \Omega_s, {\rm I})= \prod_{i=1}^N \frac{1}{S(\nu_i)}e^{-\frac{s_i}{S(\nu_i)}}
\end{equation}
where $N$ is the number of samples used in the power spectrum, $\nu_i$ is the frequency at sample $i$, $s_i$ is the observed power spectrum at frequency $\nu_i$, and $S(\nu)$ is the mean pow\-er spectrum.  In the case of single mode with no correlation with the solar background, I can write:
\begin{equation}
S(\nu)=\frac{A}{1+\left(\frac{2(\nu-\nu_0)}{\Gamma}\right)^2}+B
\end{equation}
Here I assumed, unlike \citet{Rakesh}, that the mode has no asymmetry.  Replacing Eqs. (13) and (14) in Eq. (12), the {\it posterior probability} can then be expressed as:
\begin{equation}
P(\Omega_s | {\rm D,I}) \propto P(\Omega_s | {\rm I})  \prod_{i=1}^N \frac{1}{S(\nu_i)}e^{-\frac{s_i}{S(\nu_i)}}
\end{equation}
If I assume that we know the mode amplitude $A$, the mode linewidth $\Gamma$, the noise $B$, I obtain after marginalization\footnote{We assumed that the {\it prior probabilities} are Dirac $\delta$ distributions} over these latter variables the following:
\begin{equation}
P(\nu_0 | {\rm D,I}) \propto P(\nu_0 | {\rm I})  \prod_{i=1}^N \frac{1}{S(\nu_i)}e^{-\frac{s_i}{S(\nu_i)}}
\end{equation}
This equation is almost the same as the likelihood used for fitting helioseismic power spectra by maximization.

\begin{figure*}[!]
\begin{center}
\includegraphics[width=60mm, angle=90]{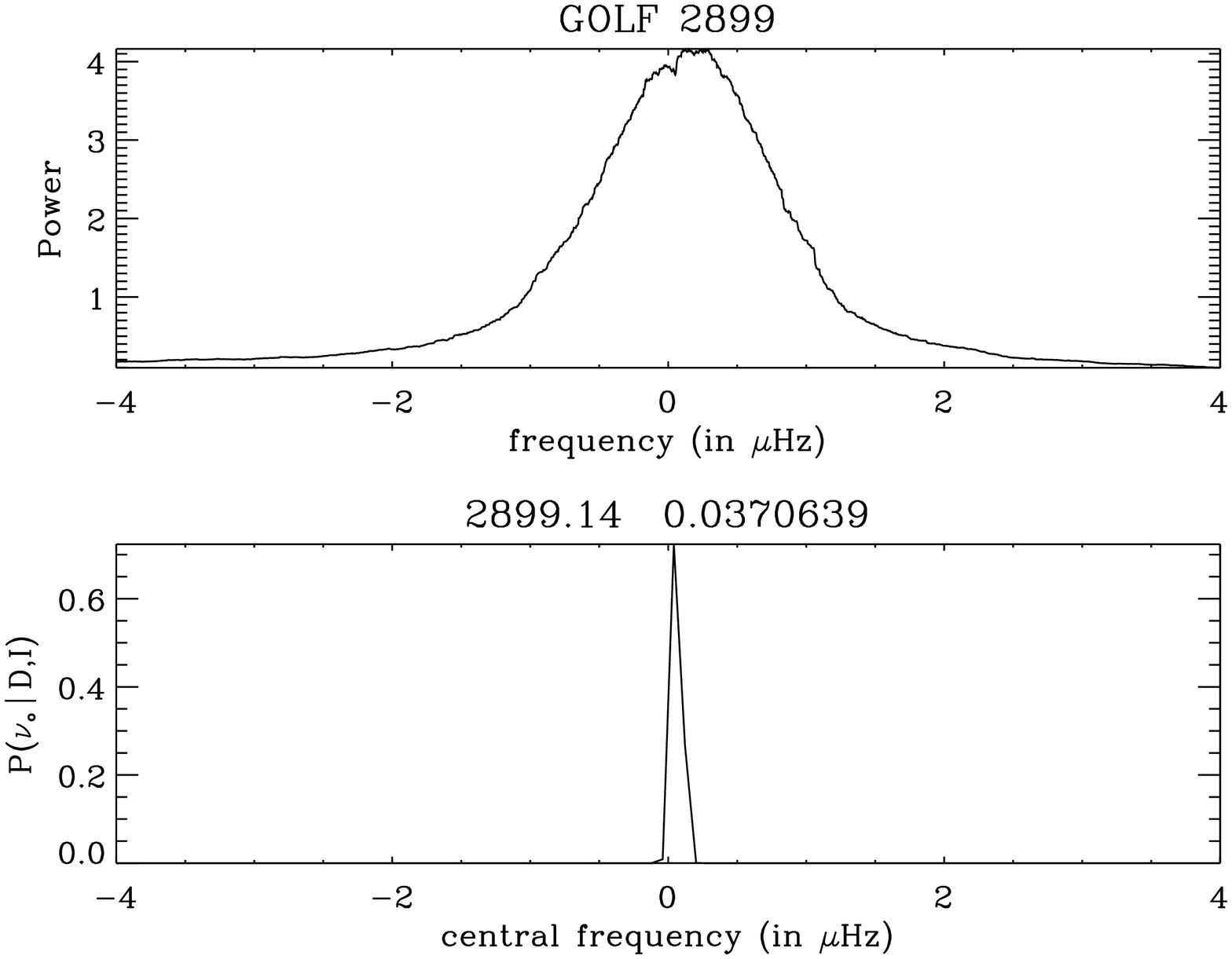}
\includegraphics[width=60mm, angle=90]{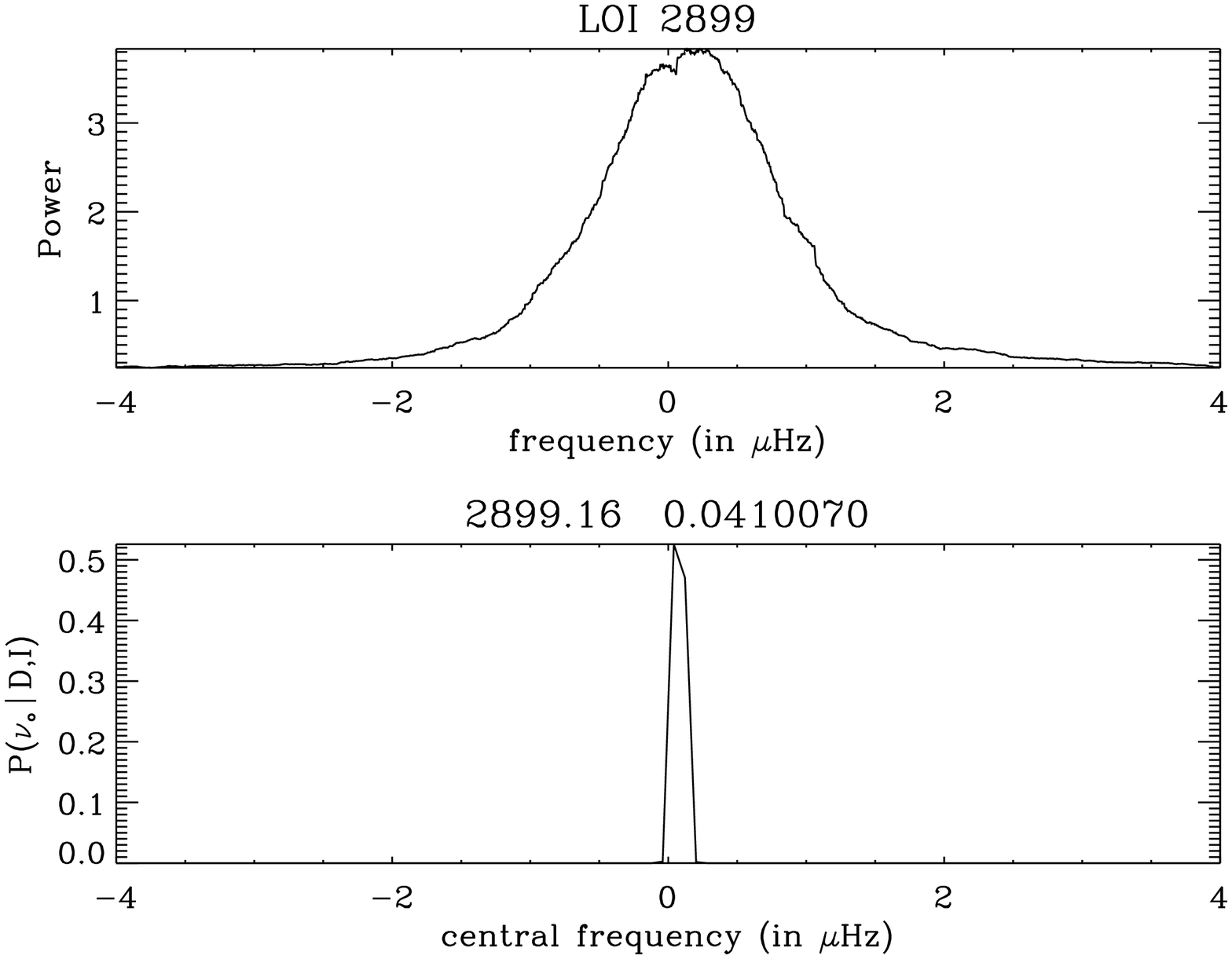}
\end{center}
\caption{(Top) Power spectra smoothed over 1$\mu$Hz (315 bins) of the GOLF and LOI spectra for an $l=0$ mode.  (Bottom) {\it Posterior probability} for GOLF and LOI data assuming an amplitude of 1 and a noise of 1, and a linewidth of 1 $\mu$Hz.  The mean and rms error of the frequency as derived from Eqs. (7) and (8) are indicated on the top of the diagrams.}
\label{label1}
\end{figure*}

\begin{figure*}
\begin{center}
\includegraphics[width=60mm, angle=90]{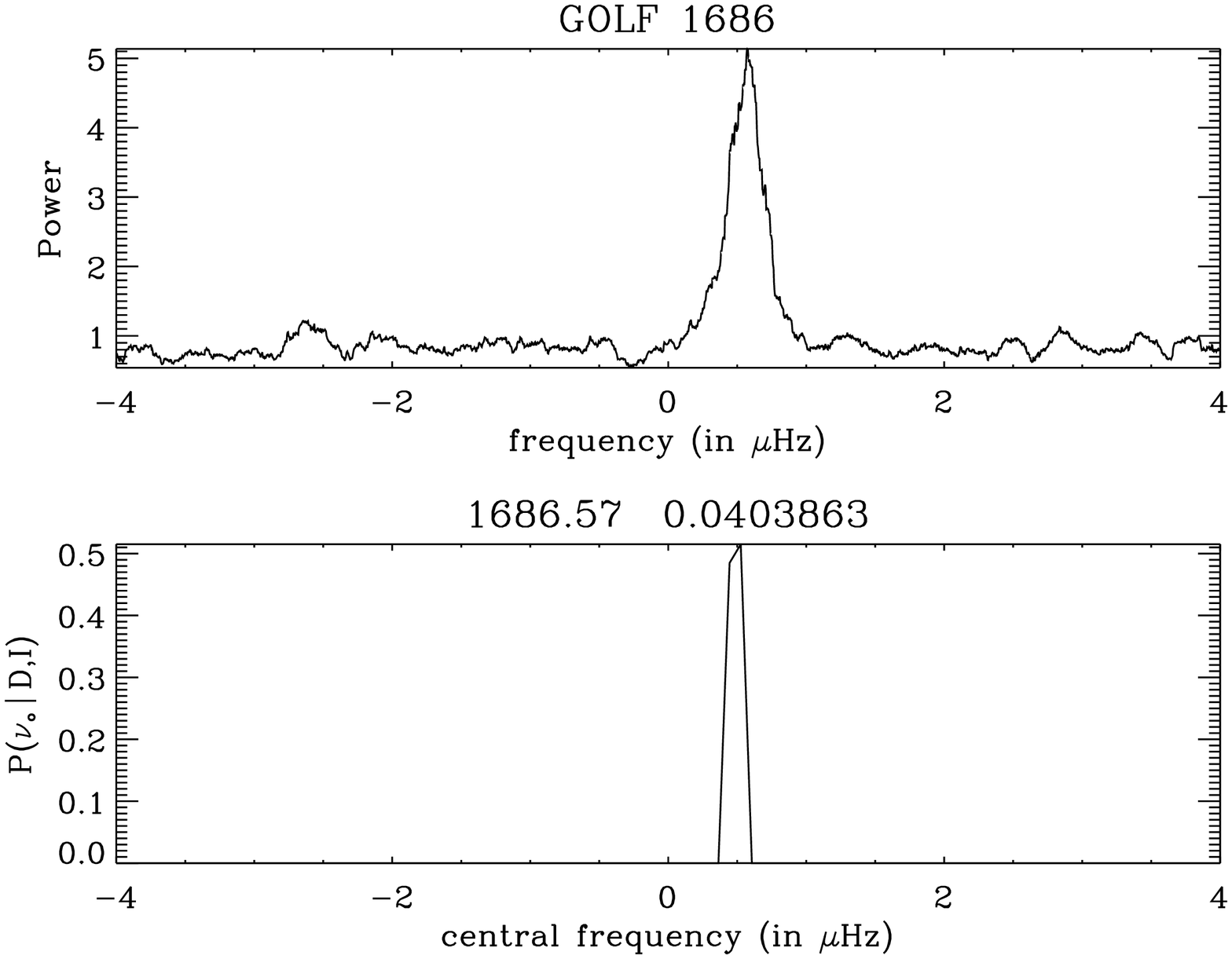}
\includegraphics[width=60mm, angle=90]{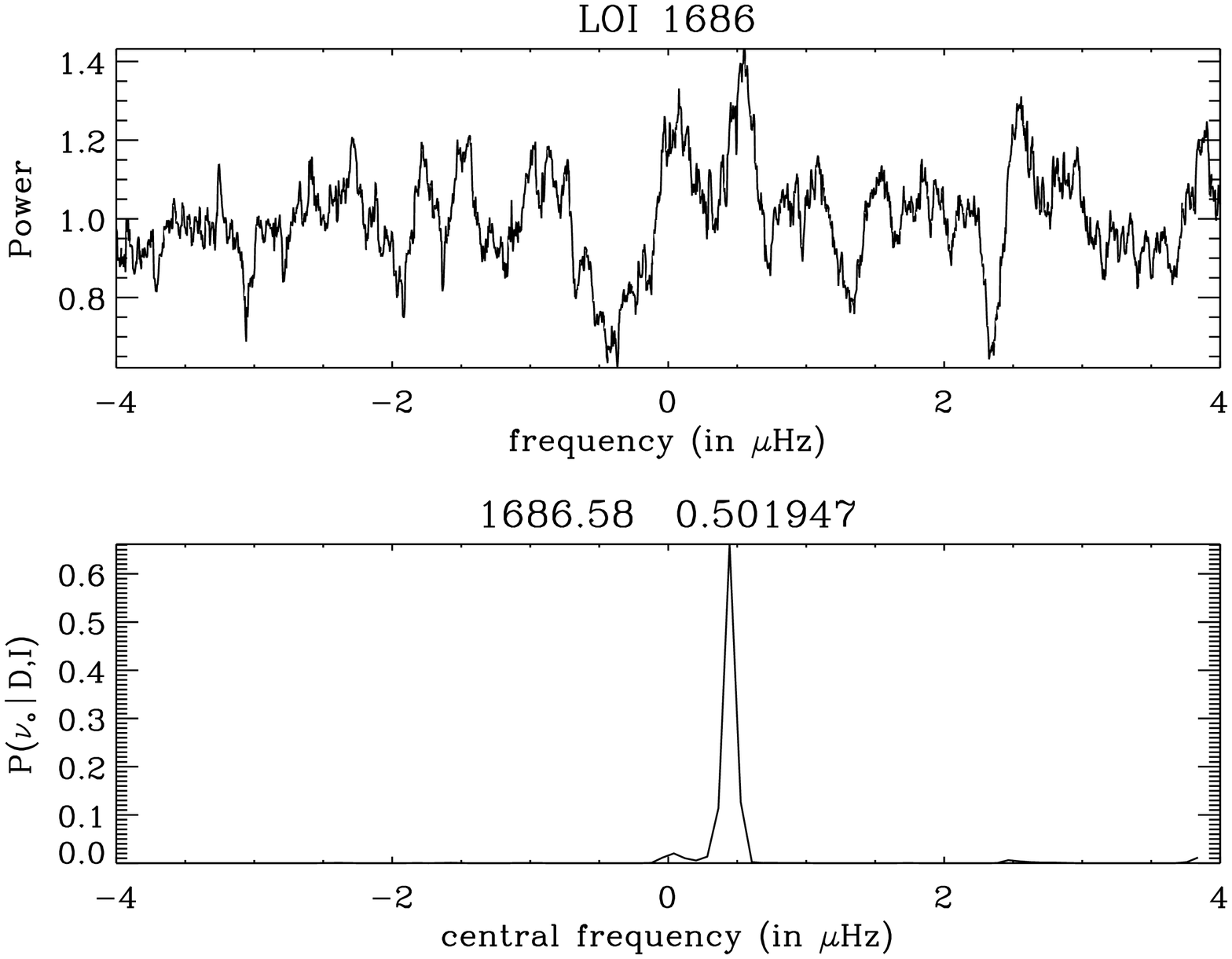}
\end{center}
\caption{(Top) Power spectra smoothed over 0.2 $\mu$Hz (63 bins) of the GOLF and LOI spectra for an $l=0$ mode.  (Bottom) {\it Posterior probability} for GOLF and LOI data assuming an amplitude of 1  a noise of 1, and a linewidth of  so 0.2 $\mu$Hz.  The mean and rms error of the frequency as derived from Eqs. (7) and (8) are indicated on the top of the diagrams.}
\label{label1}
\end{figure*}

\begin{figure*}
\begin{center}
\includegraphics[width=60mm, angle=90]{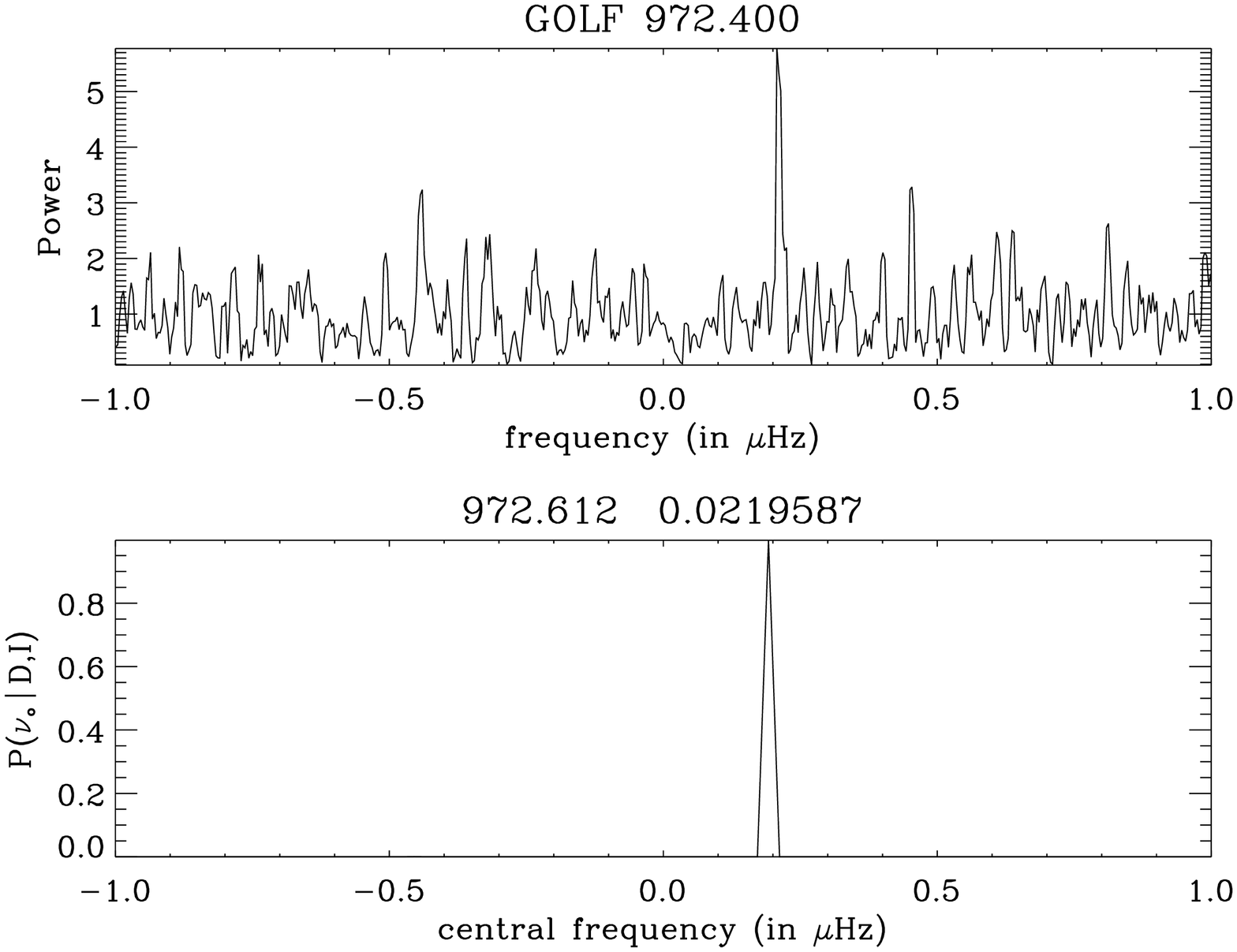}
\includegraphics[width=60mm, angle=90]{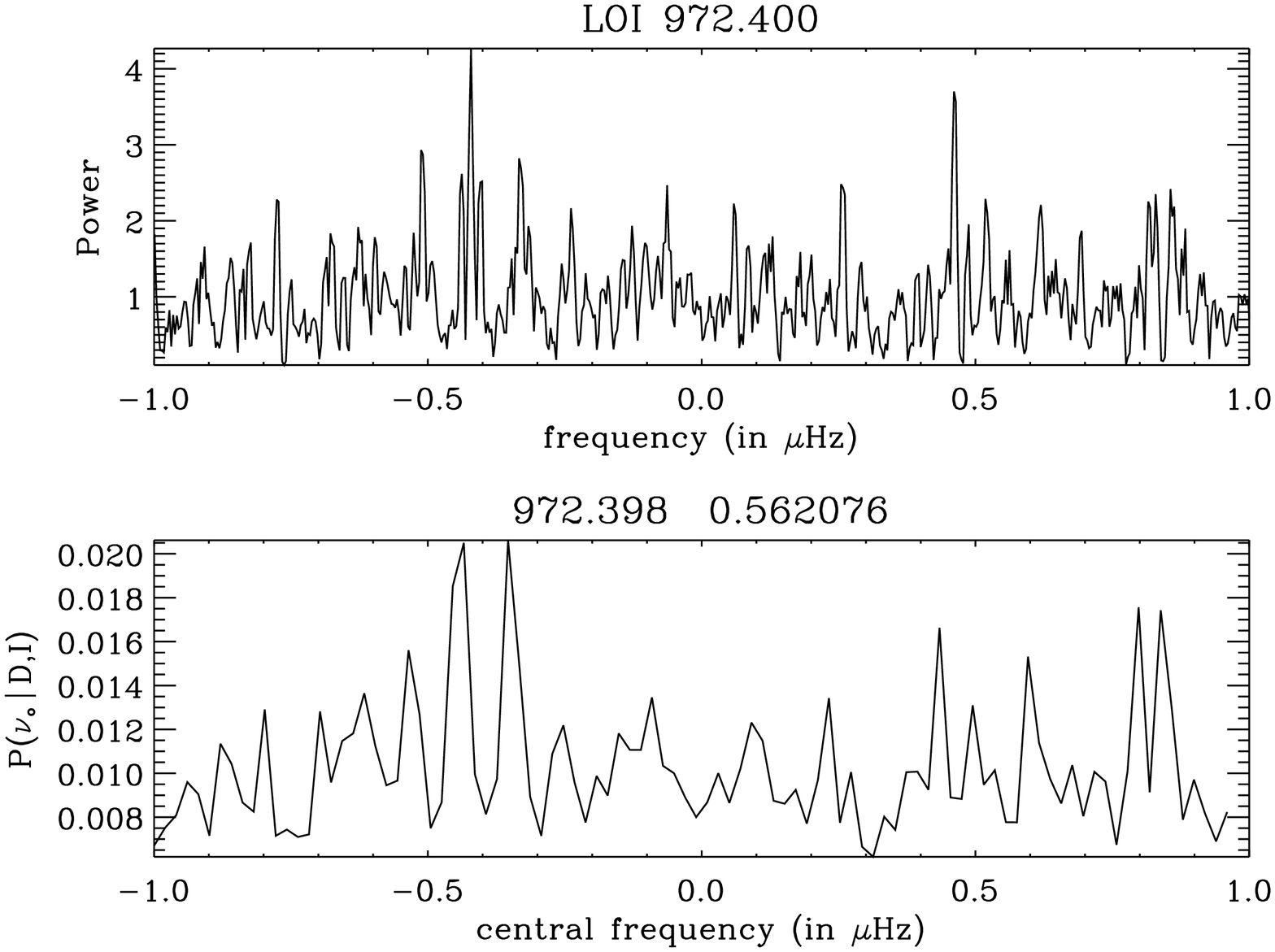}
\end{center}
\caption{(Top) Power spectra smoothed over 0.01 $\mu$Hz (13 bins) of the GOLF and LOI spectra for an $l=0$ mode.  (Bottom) {\it Posterior probability} for GOLF and LOI data assuming an amplitude of 5 and 0.2, respectively;  a noise of 1, and a linewidth 0.01 $\mu$Hz.  The mean and rms error of the frequency as derived from Eqs. (7) and (8) are indicated on the top of the diagrams.}
\label{label1}
\end{figure*}

\subsection{Applications to LOI and GOLF data}
For testing the Bayesian inference, I used almost 10 years of coeval SOHO data from the LOI\footnote{Luminosity Oscillations Imager, \cite{TABA97}} and GOLF\footnote{Global Oscillations at Low Frequency, \citet{AG97a}} instruments for testing the Bayes inference.  I applied Eq. (17) to the GOLF data for which we know modes have been detected, and to the LOI data for which modes have not been detected below 2000 $\mu$Hz.  It is well known that GOLF using solar radial velocities can detect more easily modes below 2000 $\mu$Hz.  So the Bayesian approach to the LOI data is here ideal because we know that modes exist in this frequency range, but they are almost impossible to detect directly.  In addition, solar models predict that low frequency modes are rather insensitive to surface effects.  Therefore the theoretical uncertainty provided by the model of the atmosphere is here alleviated, and the frequency of the mode is only bounded by the uncertainties in the model of the internal structure of the Sun.

Figures 1 to 3 show results for three typical cases.  Figure 1 is for a case where modes are easily detected in {\it both} instruments.  The error bars are typical of what you can also get using MLE estimators.  Figure 2 shows a very interesting case where I can detect with the LOI instrument a mode at a very low frequency close to 1500 $\mu$Hz.  In that case the error bar on the frequency is quite large.
Figure 3 shows the detection of a mode below 1000 $\mu$Hz with the GOLF instrument that was also detected by \citet{WC2002}.  On the other hand, there is no detection in the LOI because the {\it posterior probability} is now commensurate with {\it prior probability} that stated that the mode frequency was uniformly distributed over that specific range of 2 $\mu$Hz.  In that latter case, the rms error of the frequency is compatible with a frequency uniformly distributed over 2 $\mu$Hz; the rms in that case being close to $2/\sqrt{12}$.

Although the results for low frequency low degree p modes demonstrated that the Bayesian approach could be useful, the application in the g-mode range lead to results for GOLF similar to those of the LOI for the p modes (See Figure 3).  It is quite clear that additional information needs to be included before concluding for the g-mode range.  The Bayesian approach would ideally be suited to the re\-cent claims of g-mode detection made by \citet{Rafa2007}.

\section{Conclusion}
The Bayesian approach to the analysis of solar power spectra seems extremely promising.  One must not forget that many aspects have been neglected such as marginalization.  On this latter aspect, either one is fond of integration and derives the integrals of Eq. (4), or one uses the so-called Markov Chain Monte Carlo (MCMC) algorithm as used by \citet{Gregory}.  So far, I have not delved into the subject but I sense that the MCMC algorithms will need to be understood before we can proceed to the proper application of the Bayesian approach to g-mode detection.

Last but not least, this paper could be very far from a {\it proper} and {\it accurate} description of Bayesian inference.  Here I wanted to attract the {\it helioseismic} reader to a field that has recently been fast developing in astrophysics.  I hope this paper will trigger the interest of the reader for a genuine treatment of the Bayesian inference.

\acknowledgements
I benefited from useful discuss\-ions with T. Toutain, F. Baudin, N. Barbey, P. Boumier and N. Aghanim; and from a careful reading by J. Leibacher.  I also had great pleasure in reading Guy Demoment's course on {\it Mod\'elisation des incertitutes, inf\'erence logique, et traitement des donn\'ees exp\'eri\-mentales} given at Universit\'e Paris Sud - Orsay.  SOHO is a mission of international collaboration between ESA and NASA.  This paper is the result of ideas and discussions started in the framework of the International Space Science Institute whose support is greatly acknowledge.

\newpage

\bibliographystyle{aa}

\bibliography{thierrya}

\end{document}